\documentclass[journal=jacsat,manuscript=article]{achemso}
\usepackage[version=3]{mhchem} 

\author{Amy K. Lin}
\affiliation[California Institute of Technology]
{Division of Chemistry and Chemical Engineering, California Institute of Technology, Pasadena, USA}
\author{Lauren F. Illa}
\affiliation[California Institute of Technology]
{Department of Applied Physics and Materials Science, California Institute of Technology, Pasadena, USA}
\author{Natan A. Spear}
\affiliation[California Institute of Technology]
{Department of Applied Physics and Materials Science, California Institute of Technology, Pasadena, USA}
\author{Geoffrey A. Blake}
\affiliation[California Institute of Technology]
{Division of Geological and Planetary Sciences, California Institute of Technology, Pasadena, USA}
\author{Scott K. Cushing}
\email{scushing@caltech.edu}
\affiliation[California Institute of Technology]
{Division of Chemistry and Chemical Engineering, California Institute of Technology, Pasadena, USA}

\title[An \textsf{achemso} demo]
  {The Role of THz Phonons in the Ionic Conduction Mechanism of \ce{Li7La3Zr2O12} Polymorphs}

\begin{document}

\begin{abstract}
Superionic conduction in solid-state materials is governed not only by static factors, such as structure and composition, but also by dynamic interactions between the mobile ion and the crystal lattice. Specifically, the dynamics of lattice vibrations, or phonons, have attracted interest because of their hypothesized ability to facilitate fast ionic conduction. However, direct experimental measurement of the role of phonons in ionic conduction is challenging due to the fast intrinsic timescales of ion hopping and the difficulty of driving relevant phonon modes, which often lie in the low-energy THz regime. To overcome these limitations, we use laser-driven ultrafast impedance spectroscopy (LUIS). LUIS resonantly excites phonons using a THz field and probes ion hopping with picosecond time resolution. We apply LUIS to understand the dynamical role of phonons in Li\textsubscript{7}La\textsubscript{3}Zr\textsubscript{2}O\textsubscript{12} (LLZO). When in its cubic phase (c-LLZO), this garnet-type solid electrolyte has an ionic conductivity two orders of magnitude greater than its tetragonal phase (t-LLZO). c-LLZO is characterized by a disordered and vacancy-rich Li sublattice, and has a conduction mechanism dominated by single ion hops. In contrast, t-LLZO is characterized by an ordered and filled Li sublattice that necessitates synchronous ion hopping. We find that, upon excitation of phonons in the 0.5-7.5 THz range, the ionic conductivity of both polymorphs is enhanced. Furthermore, we observe a similar time decay constant of the perturbation between t-LLZO and c-LLZO despite their distinct Li sublattices and migration mechanisms. Along with Raman spectra and MD-computed vibrational density of states, these findings suggest that low-energy optical phonons perturb the lattice to enhance ionic conduction independently of the underlying static properties. We propose that it does so by increasing the population of phonons which directly modulate the ion hopping pathway and by increasing the entropy of migration thereby partially counteracting the migration barrier set by the Li sublattice. Overall, this work highlights the interplay of static and dynamic factors that enables improved ionic conductivity in both highly and poorly conducting inorganic solids alike.  
\end{abstract}

\newpage

\section{Introduction}
Superionic conductors are poised to replace organic liquid electrolytes in conventional Li-ion batteries, enabling safer all-solid-state energy storage technologies with increased energy density\cite{Tarascon2001, Armand2008, Xu2020}. While superionic conductivity can be achieved by Li\textsuperscript{+}-stuffing, tuning the vacancy concentration, and modifying structural disorder, an emerging perspective is to consider the dynamics of the crystal lattice when engineering new superionic conductors\cite{Jun2024, Muy2020}. Ionic conduction is inherently a complex process in which mobile ions interact with the surrounding framework via electrostatic and phonon interactions\cite{Tao2024}. To provide a complete picture of ionic conduction, one must consider both the dynamic and static factors that affect ion transport. The interplay of these factors is captured by the following equations which define both ionic conductivity and its prefactor: 

\begin{equation} \label{ion_cond}
\sigma_{ion} = \frac{\sigma_0}{T}e^\frac{-E_a}{k_bT}
\end{equation}
\begin{equation} \label{pre-factor}
\sigma_0 \propto e^\frac{\Delta S_m}{k_b}
\end{equation}

where $\sigma_{ion}$ is the ionic conductivity, $\sigma_0$ is the Arrhenius pre-factor, and the activation energy, $E_a$, is the experimentally measured enthalpy of migration, $\Delta H_m$. The Arrhenius pre-factor, as seen partially expanded in Equation \ref{pre-factor}, is influenced by $\Delta S_m$ representing the entropy of migration. Static factors, such as mobile ion and vacancy concentrations, are partially reflected through the enthalpy of migration, depicting the energy landscape of the system. Meanwhile, dynamic factors, such as the aforementioned phonon contributions, are captured by the entropy of migration. Notably, it has been hypothesized that raising the entropy of migration, thus increasing the Arrhenius pre-factor, can aid in offsetting large enthalpies of migration, or activation energies\cite{Muy2020}. 

Consequently, there has been significant research recently focused on how specific phonon modes affect the migration of the mobile ion. For example, it has been proposed that a lower phonon band center is associated with a reduced activation energy for ion hopping due to softening of the crystal lattice\cite{Muy2018}. Additionally, the importance of polyanion rotational phonon modes, classified as the paddlewheel mechanism, has been fiercely debated\cite{Zhang2022, Jun2024_paddlewheel}. The emphasis on phonon-ion coupling, as well as the importance of identifying low-energy optical phonon modes populated at room temperature, has been a growing trend in the field\cite{Ding2025, Gupta2024, Gordiz2021, Pham2025_THz}. 

The characterization of phonons, and specifically their effects on ionic conduction, has predominantly relied on computational methods to date, with less frequent experimental validation. In experimental studies of solid-state electrolytes, electrochemical impedance spectroscopy (EIS) is commonly used to probe ionic conductivity across multiple timescales by applying alternating current (AC) of different frequencies\cite{Wang2021}. Developments in nuclear magnetic resonance (NMR) and inelastic and quasi-elastic neutron scattering (INS and QENS, respectively) have enabled detailed experimental characterization of ion hopping by analyzing line shifts and broadening, providing time-averaged information on structural relaxation processes and mobile ion hopping rates\cite{Han2021, Schwaighofer2025}. These studies have discovered specific mechanisms by which the anionic sublattice can mediate and enhance ionic conductivity. For example, anharmonic phonons have been found to mediate a correlated hopping mechanism\cite{lin_phonon-activated_2026} and the necessity for matching timescales between polyanion reorientations and ion hops has been experimentally demonstrated\cite{hanghofer_fast_2019}. While these findings have been very impactful, none of these methods can probe site-to-site ion hopping in real time, which is crucial for capturing the rapidly evolving dynamics of the crystal lattice, nor would any of these experimental methods, in isolation, be able to correlate the population of a set of phonons with a change in ionic conductivity. To investigate the role of these phonons directly, on the inherent timescales of ion hopping, picosecond time resolution and the driving of THz-range phonon modes are required. 

\begin{figure}[hbt!]
\centering
\includegraphics[width=1\linewidth]{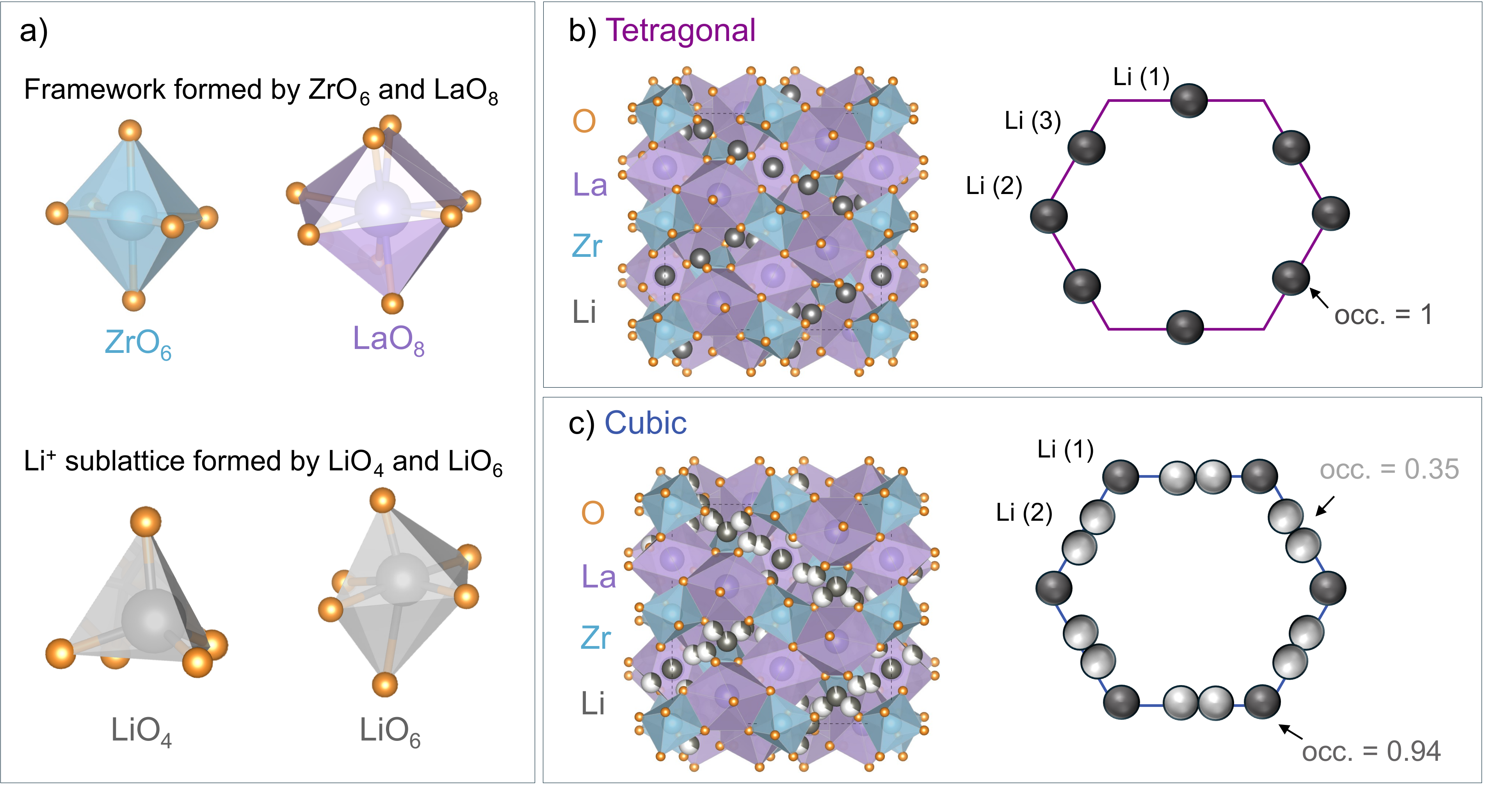}
\caption{\textbf{Polyhedra of the LLZO lattice framework and Li sublattice. Conduction pathway of (b) tetragonal and (c) cubic \ce{Li7La3Zr2O12}.} \ce{Li+} migrates across the LLZO framework of \ce{ZrO6} and \ce{LaO8} polyhedra, depicted in blue and purple, respectively\cite{Logeat2012, Chen2012}. The \ce{Li+} move through a 3D pathway characterized by individual loops connected by junctions and coordinating with O as it migrates forming \ce{LiO4} and \ce{LiO6}. An example loop for each polymorph is shown on the far right to illustrate the ordered and fully occupied Li sublattice of t-LLZO compared to the disordered and vacancy-rich Li sublattice of c-LLZO\cite{Awaka2009, Awaka2011}.}
\label{mt_fig1}
\end{figure}

Accordingly, we use laser-driven ultrafast impedance spectroscopy (LUIS) in this work, as described in previous articles\cite{Pham2024, Pham2025_THz}, and discussed in detail in the next section. Herein, we experimentally study the interplay of static and dynamic properties and its effects on ionic conduction. Specifically, we focus on the garnet-type Li\textsuperscript{+} conductor, Li\textsubscript{7}La\textsubscript{3}Zr\textsubscript{2}O\textsubscript{12} (LLZO) as a case study material. LLZO has garnered attention due to its low electronic conductivity, thermal and chemical stability, and wide electrochemical stability window, all of which are conditions necessary for use in a solid-state battery\cite{Kravchyk2022, Chen2020}. Importantly, the cubic polymorph of LLZO (c-LLZO) has an ionic conductivity (0.1-1 mS/cm) approaching that of liquid electrolytes (10 mS/cm)\cite{Murugan2007}. However, c-LLZO is only formed above $\sim$400-$\sim$600 $^{\circ}$C, whereas its tetragonal polymorph (t-LLZO) is the stable phase at room temperature \cite{Gullbrekken2025, Jung2022}. In both polymorphs, it is computationally postulated that Li\textsuperscript{+} moves through LLZO's framework of ZrO\textsubscript{6} and LaO\textsubscript{8} polyhedra via junction-connected loops of LiO\textsubscript{4} tetrahedra and LiO\textsubscript{6} octahedra as illustrated in Figure \ref{mt_fig1}\cite{Awaka2011, Meier2014}. 

Despite their matching frameworks, the ionic conductivity of t-LLZO is two orders of magnitude lower than that of c-LLZO. This is often attributed to differences in the Li sublattices of the two polymorphs\cite{Awaka2009, Meier2014}. T-LLZO has three Li sites, all fully occupied, whereas c-LLZO has two Li sites with reduced occupancy\cite{Awaka2009, Awaka2011}. The predicted occupancy values from X-ray diffraction analysis are g = 0.94 and g = 0.35 at the Li1 and Li2 sites, respectively\cite{Awaka2011}. Neutron diffraction studies report an even lower occupancy of g = 0.56 and g = 0.44\cite{Xie2011}. Accordingly, due to this disordered and vacancy-rich Li sublattice which enables cooperative single-ion hops, c-LLZO has a low ion hop activation energy of $\sim$0.3 eV\cite{Sakamoto2013, Meier2014}. Conversely, t-LLZO has a high activation energy of $\sim$0.5 eV because synchronous ion hopping is required to move through its ordered and filled (g = 1) Li sublattice\cite{Awaka2009, Awaka2011, Meier2014}. Beyond this static picture, recent research suggests the presence of THz-range phonons governs the \ce{Li+} migration mechanism in c-LLZO, specifically through lattice softening\cite{Kim2025}. As such, the LLZO polymorphs, with their identical frameworks yet differing Li sublattices, offer a unique opportunity to probe the effects of dynamically perturbing the same lattice phonons in two material systems with quite different static properties.

To this end, we use LUIS to resonantly excite low-energy optical phonon modes from 0.5-7.5 THz, where modes populated at room temperature reside, and subsequently probe changes in ion hopping in both c-LLZO and t-LLZO. Upon resonant excitation of these THz-frequency phonon modes, we find an enhancement in the ionic conductivity of both c-LLZO and t-LLZO. We notably find that the perturbation decays out in both polymorphs in approximately one ns despite their differing Li sublattices and migration mechanisms. This suggests that the low-energy optical phonons highlighted in this investigation alter the lattice framework and mediate ionic conduction in a manner independent of the underlying properties of the Li sublattice. Such phonon-ion interactions are inherently transient and may have eluded detection through time-averaged or static studies of the crystal lattice. As such, the continued investigation of these many-body interactions which occur during site-to-site hopping in real time may allow for the emergence of new design rules which can be derived from phonon-ion correlations.

\section{Instrumental Set-Up}

\begin{figure}[hbt!]
\centering
\includegraphics[width=0.6\linewidth]{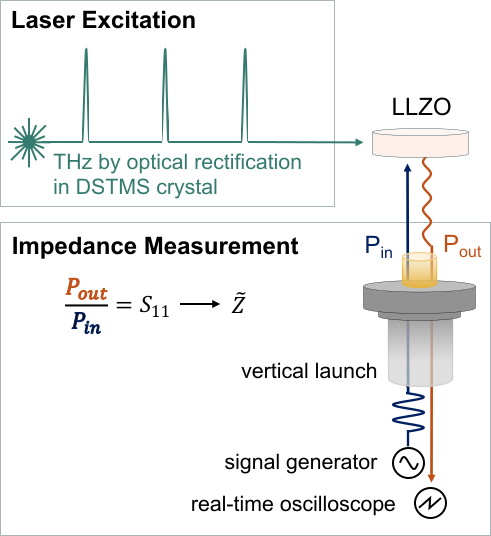}
\caption{\textbf{Laser-driven ultrafast impedance spectroscopy (LUIS).} Broadband 0.5-7.5 THz radiation is directed toward a 300 $\mu$m thick and 1 mm wide LLZO pellet. Orthogonally, a GHz AC waveform, shown in blue, is transmitted to the pellet from a signal generator via a high frequency coaxial cable, a vertical launch, and, finally, a gold pin which interfaces with the pellet. Impedance mismatch at that interface, in an otherwise impedance-matched system, causes a portion of the GHz signal, shown in orange, to be reflected back through the high frequency coaxial cable and is sampled by a real-time oscilloscope. When the THz excitation causes a change in the LLZO's impedance, the perturbed GHz signal back-reflection is captured by the real-time oscilloscope.}
\label{mt_fig2}
\end{figure}

In LUIS, an ultrafast laser pulse is used to excite lattice dynamics of interest in a material, in this work c-LLZO or t-LLZO. The corresponding change in the ionic conductivity of the LLZO polymorph is probed using a high frequency alternating current (AC) wave. The method is depicted schematically in Figure \ref{mt_fig2}. In this work, the samples of interest are pressed pellets of either c-LLZO or t-LLZO. The pellets are prepared to be 300 $\mu$m thick and 1 mm in diameter such that they can rest atop a gold pin. This gold pin is connected to a high frequency coaxial cable through which a 19 gigahertz (GHz) and -5 dBm (172 mV sinus amplitude) AC wave is transmitted from a signal generator to the LLZO. The GHz signal travels through a fully impedance-matched circuit such that all AC waveforms are transmitted with minimal loss and the only impedance mismatch occurs at the LLZO-pin interface. This impedance mismatch forces part of the AC waveform to be reflected and is routed to a real-time oscilloscope. The latter captures changes in impedance by recording the reflected waveform every 7.8 ps. This methodology is described in detail in several previous works\cite{Pham2024,Pham2025_THz}.

As discussed previously, in this work we are interested in measuring the role of low-energy optical phonon modes, particularly those that lie at or below $\sim$6 THz, meaning they are populated at room temperature. To excite these phonon modes, we generate THz light with frequency content spanning 0.5-7.5 THz and a field strength of 744 kV/cm using optical rectification in a DSTMS (4-N,N-dimethylamino-4'-N'-methyl-stilbazolium 2,4,6-trimethylbenzenesulfonate) crystal (Figure S2). This light is used to excite the sample, which leads to a time-domain perturbation within the otherwise impedance-mismatched reflected waveform. We note that the perturbed signal only represents the effects of the phonon mode at the gamma point on impedance.

Isolating this perturbed signal within the GHz AC carrier wave is quite challenging for several reasons. First, there is inherent noise associated with high frequency measurements. Second, we have identified a long-range antenna pick-up that is present even when the THz light is blocked, as seen in Figure S5. This signal is not trivial, but notably different from the cases where the sample is directly excited. Finally, we note the differential nature of the technique. The goal of the measurement is not to maximally enhance ionic conductivity, but instead to use THz excitations to gain a deeper insight into phonon-mediated ionic conduction. Considering these challenges, we are still able to extract an informative signal, discussed later on, but we also propose several technical advancements that should be considered in the future. Future techniques aimed at capturing these dynamical process should enhance the signal-to-noise ratio, improve sample isolation to reduce any antenna-like pick-up, and allow for the simultaneous measurement of absolute and differential changes to gain a more complete understanding of ionic conduction.

Given the current limitations, we apply significant averaging and post-processing. As we record the measurement, the oscilloscope averages the incoming waveform 1024 times. Subsequently, an amplitude demodulation function is applied, and the resulting demodulated signal is further averaged 1024 times. Finally, the moving root-mean-square average of the time-domain spectrum is computed. The final signal is fitted with an exponentially modified Gaussian function, as described in the Methods section, in which the Gaussian and exponential components capture the rise and decay of the perturbation, respectively.

\section{Results \& Discussion}

\begin{figure}[hbt!]
\centering
\includegraphics[width=0.75 \linewidth]{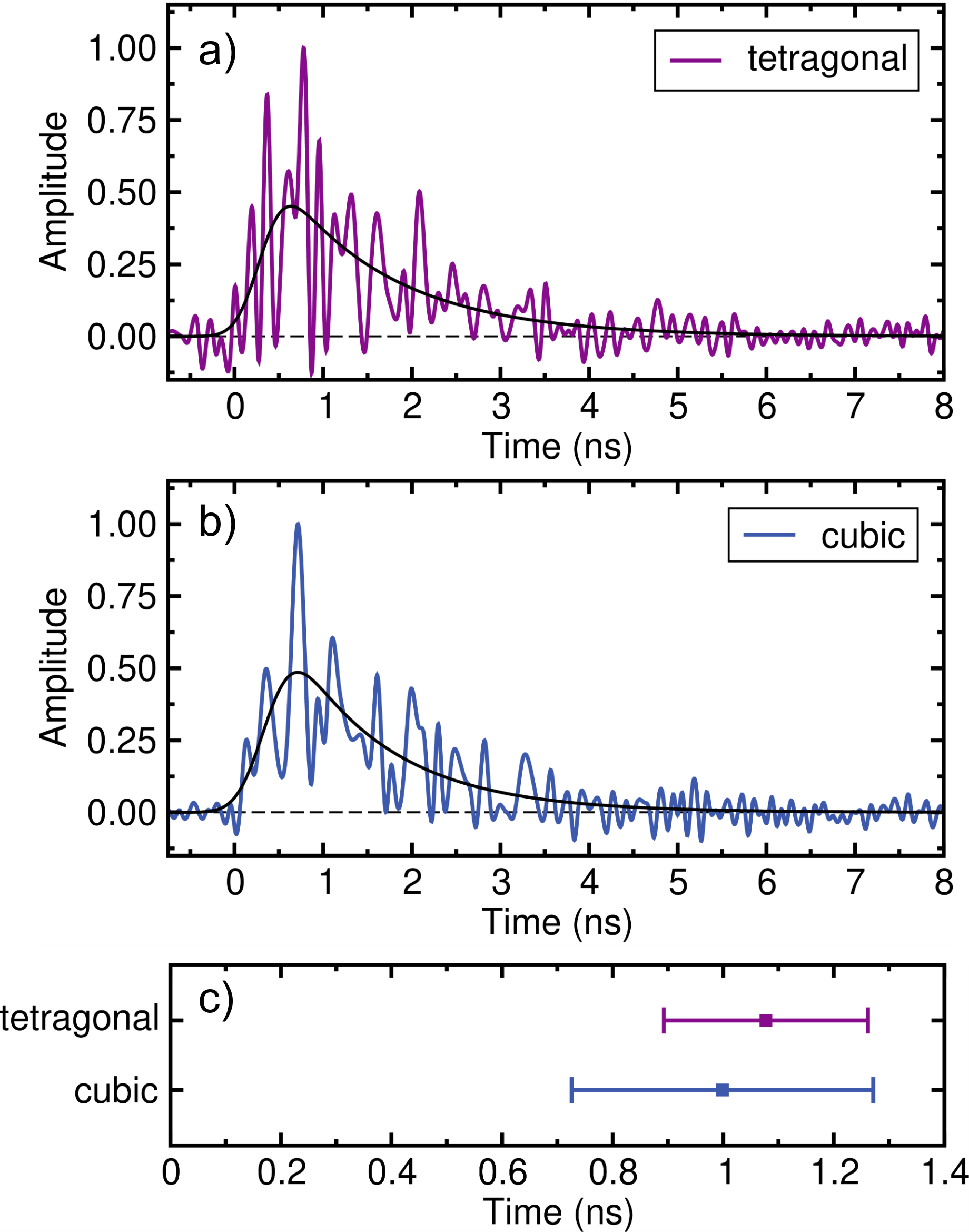}
\caption{\textbf{Time-resolved THz-field-induced perturbations in the impedances of (a) t-LLZO and (b) c-LLZO with their average decay time constants across 12 replicates reported in (c).} The rise and decay of a perturbation in the reflected GHz signal by t-LLZO (a) and c-LLZO (b), represent their change in impedance as a result of resonantly exciting phonons between 0.5-7.5 THz by a THz field (744 kV/cm). The purple and blue signals are the root-mean-square averages of amplitude-demodulated spectra. The solid black lines are the exponentially modified Gaussian fit of each spectrum from which the decay time constant is extracted. The spectra have been shifted vertically so that the center of the GHz carrier is at zero, and horizontally so that time zero corresponds to the THz impulse. Across 12 replicates each, we find the average decay time constant to be $\sim$1.08 ns and $\sim$1.00 ns for t-LLZO and c-LLZO, respectively, represented in (c).}
\label{mt_fig3}
\end{figure}

Figure \ref{mt_fig3} shows a perturbation in the reflected waveform by t-LLZO and c-LLZO as a result of the THz-range excitation pulse. It is worth noting that the c-LLZO used in this experiment is stabilized at room temperature by Nb-doping primarily at the Zr site (Li\textsubscript{6.5}La\textsubscript{3}Zr\textsubscript{1.5}Nb\textsubscript{0.5}O\textsubscript{12}). X-ray diffraction of t-LLZO and c-LLZO, confirming the purity of the samples, can be found in Figure S1. The decay time constant extracted from the exponentially modified Gaussian fit averaged across 12 replicate measurements of each polymorph is 1.08 ns $\pm$ 0.19 ns and 1.00 ns $\pm$ 0.27 ns for t-LLZO and c-LLZO, respectively. 

The ns decay following the THz-range excitation is contrasted by the rapid decay of the perturbation due to below band gap excitations by a 400 nm ($\sim$12 $\mu$J/pulse) or 800 nm ($\sim$8 $\mu$J/pulse) femtosecond laser, as shown in Figures S3 and S4, respectively. The decay time constants under both excitations are below 30 ps for both the t-LLZO and c-LLZO. Since the band gap of LLZO is approximately 6 eV (206 nm), the 400 nm and 800 nm pulses do not drive band-edge excitations\cite{Thompson2017} and previous work has shown that such excitations behave similarly to laser-induced heating, such as the decay of acoustic phonons\cite{Pham2025_UV}. The difference in the length of the perturbations induced by the THz impulse and those induced by the 400 nm and 800 nm impulses implies that the THz excitation drives lattice dynamics distinct from laser-induced heating. The ns-long relaxation might suggest that the signal describes ion transport dynamics which are generally slower than phonon-specific decay pathways which would be expected to occur on sub-ns timescales. We can also rule out the contributions of electrons to the measured impedance perturbation due to the low electronic conductivity of LLZO, $10^{-8}$ S/cm, the large band gap of the material (6 eV), and previous UV excitation experiments\cite{Kravchyk2022, Thompson2017, Pham2025_UV}. For all these reasons, we hypothesize that the excitation energy injected into the system induces a measured change in impedance that specifically reflects increased ion hopping in the LLZO polymorphs. 

Strikingly, the decay time constant of t-LLZO's LUIS response is observed to be indistinguishable from that of c-LLZO, as indicated by their overlapping standard deviations (Figure \ref{mt_fig3}c). The $\sim$1 ns decay observed in both polymorphs is an intriguing result since one might hypothesize that a difference in relaxation time should arise from their distinct Li\textsuperscript{+} migration processes. In c-LLZO, although Li\textsuperscript{+} can migrate cooperatively, its migration is dominated by single hops in a vacancy-rich disordered sublattice, which allows for an increased number of stable ion configurations compared to t-LLZO. Accordingly, this should facilitate the ions' rapid return to an equilibrium state characterized by a faster decay of the impedance signal. Along these lines, it would be expected that the necessary synchronous movement of Li\textsuperscript{+} through t-LLZO's packed and ordered Li sublattice would induce a longer return to equilibrium. However, the striking similarity between the two signals suggests that the laser-excited THz-range phonons affect conduction in a manner irrespective of the properties of the Li sublattices and migration mechanisms. In line with these results, we propose a new hypothesis: THz-range phonons seem to enhance the impedance of both LLZO polymorphs by perturbing the structural framework shared by the two materials, resulting in their similar ultrafast responses and decay times.    

\begin{figure}[hbt!]
\centering
\includegraphics[width=0.75 \linewidth]{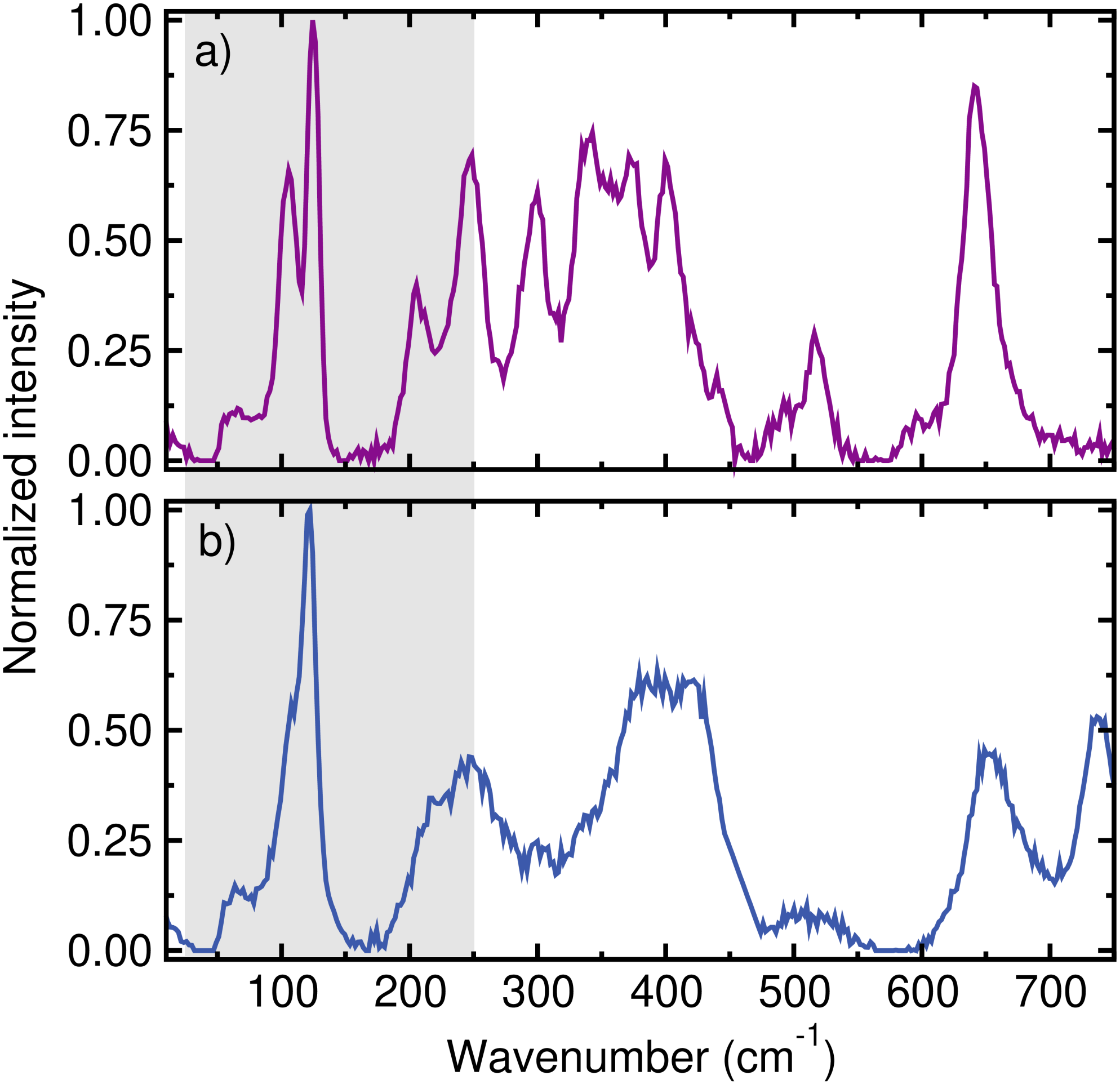}
\caption{\textbf{Raman spectra of (a) t-LLZO and (b) c-LLZO.} The Raman-active modes in the LUIS excitation region of 0.5-7.5 THz ($\sim$17-250 cm\textsuperscript{-1}) are highlighted in gray. The modes in the emphasized region can largely be attributed to La and Zr structural modes associated with the shared framework of the LLZO polymorphs and are sharp reflecting its rigidity. Notably, this LUIS measurement does not excite the mid-frequency region dominated by Li-O character where c-LLZO and t-LLZO differ as evidenced by their broad and sharp peaks, respectively, representing the nature of their Li sublattices.}
\label{mt_fig4}
\end{figure}

\begin{figure}[hbt!]
\centering
\includegraphics[width=0.75 \linewidth]{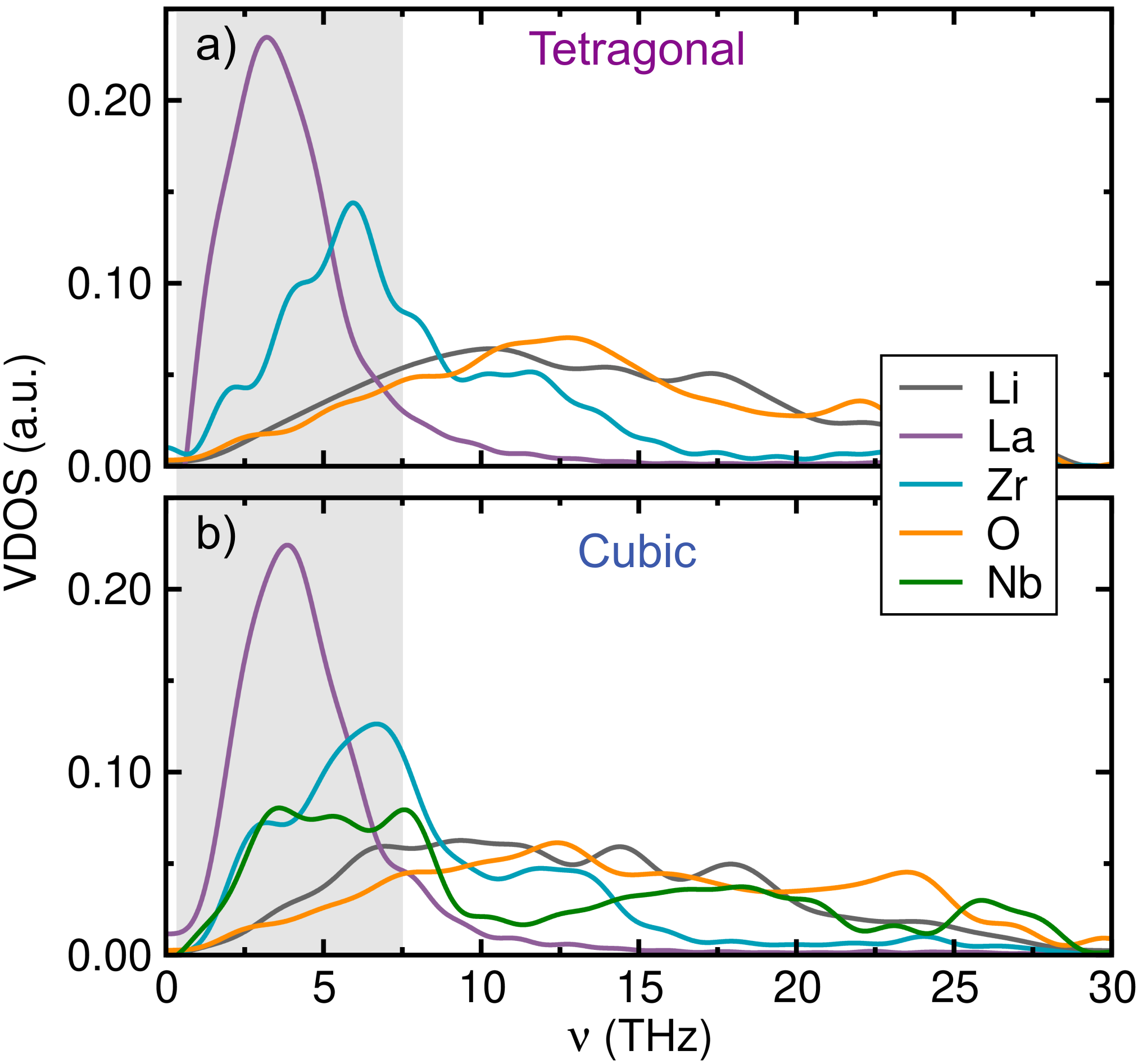}
\caption{\textbf{Vibrational density of states spectra of (a) t-LLZO and (b) c-LLZO.} The VDOS is computed using classical MD and is a representation of all modes, not only those which are optically accessible. The region coherently driven by the LUIS THz laser excitation is shaded in gray and dominated by the framework's La and Zr character. Strikingly, the spectra of both polymorphs are very similar in this region. These observations support the claim that the THz excitation targets phonons associated with the shared framework of c-LLZO and t-LLZO.}
\label{mt_fig5}
\end{figure}

To better evaluate such a proposal, the t-LLZO and c-LLZO phonon baths are considered. As expected, in the Raman spectra, Figure \ref{mt_fig4}, we observe broader peaks in c-LLZO than t-LLZO within the mid-frequency region of 200-500 cm\textsuperscript{-1} associated with Li-O\cite{Tietz2012}. This reflects the fluctuating Li\textsuperscript{+} coordination environments found in the more disordered energy landscape of the Li sublattice in c-LLZO. In the low-frequency THz regime where the LUIS THz excitation occurs, one can find structural modes associated with the framework of La, Zr, and dopant vibrations\cite{Kim2025}. Particularly, the region below 200 cm\textsuperscript{-1} is dominated by La vibrations and includes the main Zr peak whose character extends out to 400 cm\textsuperscript{-1}\cite{Tietz2012}. Furthermore, both polymorphs have a rigid host lattice as evidenced by the sharp and well-resolved peaks associated with the framework, found in the $<$200 cm\textsuperscript{-1} region and with the Zr-O peak at $\sim$640 cm\textsuperscript{-1}. Critically, as illustrated in the gray-shaded region of Figure \ref{mt_fig4}, c-LLZO and t-LLZO have relatively comparable phonon baths in the low-energy THz regime where the LUIS excitation occurs. When we extend this discussion to the analysis of the vibrational density of states (vDOS) of the two polymorphs, a consistent story emerges. The vDOS, computed using classical molecular dynamics (MD) and described in detail in the Methods sections, allows for the identification of the full vibrational bath, not simply modes which are Raman-active or optically accessible. The gray-shaded region in Figure \ref{mt_fig5} represents the energy-range accessed by the LUIS THz excitation. This region is very similar in both t-LLZO and c-LLZO and is largely dominated by the framework's La and Zr character. This evidence supports the data from the transients presented in Figure \ref{mt_fig3} depicting the indistinguishable result of coherently driving phonons associated with the shared lattice framework in these polymorphs in this energy range.

We propose two ways in which the structural framework of LLZO (Figure \ref{mt_fig1}) can modulate Li\textsuperscript{+} migration independent of the existing Li sublattice. First, the oxygen atoms of LaO\textsubscript{8} and ZrO\textsubscript{6} define the area of the structural bottlenecks through which Li\textsuperscript{+} must pass during a hop. Since the THz laser excitation distorts these polyhedra, we hypothesize that the bottleneck is modulated, and with it, the migration barrier. Second, because the LiO\textsubscript{4} and LiO\textsubscript{6} polyhedra of the Li\textsuperscript{+} migration pathway share oxygens with the framework, perturbations of the La-O and Zr-O bonds should alter the strength of Li-O coordination. This changes how strongly Li\textsuperscript{+} is bound to its site and thus, its propensity to hop to the next one. Both mechanisms initiated by the THz excitation would affect the two polymorphs similarly regardless of their Li sublattice characteristics because of their shared frameworks. Moreover, LUIS is a differential measurement characterizing impedance relative to its unperturbed state, thus the static distinctions of each polymorph are not captured by this measurement. For these reasons, a comparable enhancement in the ionic conductivity and subsequent decay behavior in both polymorphs is observed.  

The connection between enhanced ionic conductivity and framework-associated phonon modes is further supported by previous studies of another solid-state electrolyte, Li\textsubscript{3x}La\textsubscript{2/3-x}TiO\textsubscript{3} (LLTO). In LLTO, lattice deformation of the structural 4-O bottleneck, caused by the tilting of \ce{TiO6} octahedra, facilitates \ce{Li+} migration via a concerted diffusion mechanism\cite{Gao2025}. In another piece of computational work, it is shown that these \ce{TiO6} tilting modes largely reside in the region between 1-5 THz and that ionic conductivity in LLTO is enhanced upon resonant excitation of these modes\cite{Pham2025_THz}. Analogously, irrespective of the mechanism of the mobile ion, the 0.5-7.5 THz modes in LLZO can modulate the framework to facilitate enhanced ionic conductivity in both polymorphs. Following this logic, it is possible that the enhancement in ion hopping is brought about by targeted excitation of specific phonon modes which couple with ion hopping.

Still, the comparable enhancements inform us about the differing roles of phonon-ion coupling in the two polymorphs. In the case of c-LLZO, the notably similar enhancements suggest that its superior ionic conductivity is overwhelmingly attributable to its static properties, namely, its vacancy-rich, disordered Li sublattice. Alternatively, the similar enhancements in the case of t-LLZO suggest that increasing the population of low-energy optical phonon modes which mediate the hopping bottleneck could be key to enhancing ionic conductivity in otherwise poorly conducting inorganic solids. In total, we suggest that increasing the population of THz-range phonon modes associated with the crystal framework can modulate, and even enhance, ionic conductivity independent of the mobile ion sublattice. In addition to exciting specific bottleneck modes, the THz excitation may also increase the entropy of migration which can offset some of the activation energy effects determined by the energy landscape of the mobile ion sublattice, a static factor. 

In detail, the entropy of migration encompasses both configurational and vibrational entropies. Configurational entropy is determined by static properties governing the number of ways in which the atoms in a system can be arranged, independent of temperature\cite{Fultz2020}. As such, we hypothesize that we are altering the vibrational entropy. As the THz-range phonons are coherently driven, analogous to thermally populating select phonons, a larger volume of phase space can be explored by the framework's atoms\cite{Fultz2020}. This augments the population of vibrational states in the system thereby increasing vibrational entropy. As the vibrational entropy of the framework elevates, this may increase the difference in vibrational entropy between the equilibrium state and the saddle point (i.e. transition state) which is represented by the entropy of migration. Accordingly, we propose that targeting materials with an enhanced population of phonons in this regime can increase the entropy of migration, thereby raising the Arrhenius pre-factor, in line with the Meyer-Neldel or compensation rule\cite{Muy2020}. Combined with our previous discussion, when these modes specifically alter the ion hop pathway, this can create a synergistic effect, whereby the enhanced ionic conductivity can be further amplified.

\section{Conclusion}
The tetragonal and cubic polymorphs of LLZO are a compelling pair of materials to study because, despite the similar \ce{ZrO6} and \ce{LaO8} framework and stoichiometry of these polymorphs, their ionic conductivities differ by two orders of magnitude. While the pair has been investigated using a static lattice representation, this work aims to derive design rules from dynamic lattice measurements by examining how the population of THz-range optical phonons, which are populated at room temperature, affect their ionic conductivities. Using a THz pulse to resonantly excite phonons in both polymorphs, the impact of these vibrations on site-to-site \ce{Li+} conduction is characterized in real-time using LUIS. A differential enhancement in ionic conductivity is observed for both polymorphs where the measured decay of the impedance signal, which informs on the ion migration dynamics, is found to be unexpectedly similar in both materials despite their differing Li sublattices and migration mechanisms. Together, with Raman spectra and MD-computed vibrational density of states, this result suggests that the excited THz-range phonons perturb the LLZO framework in the same manner irrespective of the static properties of the polymorphs. We hypothesize that this is likely due to increasing the entropy of migration and partially offsetting the enthalpy of migration set by the Li sublattice and migration mechanism. In total, this case study indicates the importance of understanding how a dynamical lattice can enhance ionic conductivity in both promising and poor solid-state ionic conductors, while also emphasizing the importance of the interplay between static and dynamic factors in the engineering of highly ion-conducting solids.

\section{Methods}

\textbf{LLZO Synthesis}
t-LLZO is synthesized according to literature\cite{Awaka2009}. Stoichiometric amounts of \ce{La2O3}, \ce{ZrO2}, and 10 \% weight excess of \ce{Li2CO3} (accounts for Li loss during calcination), are mixed with a mortar and pestle. Pellets are made from the mixture using a manual Arbor press. The pellets are placed on a crucible with sacrificial powder both underneath and on its surface. The calcination is done in a furnace at 900 $^\circ$C for 5 hours at a ramp rate of 1 $^\circ$C/minute.

Polycrystalline Nb-doped c-LLZO is made by grinding single crystals of the material, sent from the National Institute of Advanced Industrial Science and Technology in Japan, using a mortar and pestle. 

Both t-LLZO and c-LLZO are annealed according to literature protocols\cite{Awaka2009}. After calcination, the t-LLZO is ground into a powder using a mortar and pestle. Both the t-LLZO and c-LLZO powders are made into 6 mm wide, 1 mm thick pellets by pressing under 2 tons of pressure for two 7 minute periods with a hydraulic press. Then, the pellets are placed on a crucible and covered in sacrificial powder. The anneal step is done at 980 $^\circ$C for 5 hours at a ramp rate of 1 $^\circ$C/minute. Lastly, the pellets are shaved down with sandpaper to be 1 mm wide and 300 $\mu$m thick. 

\textbf{X-ray Diffraction of LLZO}
X-ray diffraction of t-LLZO and c-LLZO is obtained using a Rigaku X-ray diffractometer with CuK$\alpha$ radiation (Figure S1). The measurement is performed with a knife edge to minimize background and the LLZO powders are scanned from 2$\theta$ = 10$^\circ$ to 70$^\circ$ with a scan rate of 0.1$^\circ$/minute. Rietveld refinement is done on the resultant spectra with GSAS II software.  

\textbf{Raman Spectroscopy of LLZO}
Raman spectra of t-LLZO and c-LLZO powders are obtained using a Renishaw inVia QONTOR micro-Raman Spectrometer System with a 514.5 nm laser. The presented spectrum is the result of 200 average spectra, each taken with an acquisition time of 1 second, at 10\% average power, with a 50 $\mu$m slit and a 500 $\mu$m aperture.

\textbf{Excitation Sources}
\newline
800 nm: A Ti:sapphire laser is used to generate 38 femtosecond pulses of 800 nm light at a repetition rate of 1 kHz. Neutral density filters are used to bring the energy of the pulse, which excites the LLZO, to around 8 $\mu$J/pulse. 

400 nm: The 800 nm light from the Ti:sapphire laser is used to pump a beta-barium borate nonlinear crystal to produce 400 nm light through second harmonic generation. Neutral density filters are used to bring the energy of the 400 nm pulse, which excites the LLZO, to around 12 $\mu$J/pulse.

THz generation: The 800 nm light from the Ti:sapphire laser is sent through an optical parametric amplifier to produce 1300 nm light with a repetition rate of 500 Hz. A DSTMS crystal is pumped by the 1300 nm light, and through optical rectification in the nonlinear crystal, a broadband THz field is generated with a field strength of 744 kV/cm and a bandwidth of approximately 0.5-7.5 THz. Due to high absorption of THz frequencies by air, the THz field is maintained in a purge box back-filled with nitrogen gas.   

THz detection: The THz frequency content is detected through an electro-optic sampling scheme. The P-polarized THz field passes through and thereby induces a birefringence in a GaP THz detector crystal. Linearly polarized 800 nm light from the Ti:sapphire laser hits the now birefringent GaP crystal, which rotates the plane of polarization of the 800 nm light. The arrival time of the 800 nm light is scanned such that shifts in the 800 nm polarization can be detected as a function of time by balanced photodiodes to generate a time-domain spectrum (Figure S2). Subsequently, via a fast Fourier transform of the time-domain spectrum, the frequency-domain spectrum of the generated THz field is obtained (Figure S2).   

\textbf{Decay Time Constant Fitting}
The spectra obtained through THz-excitation LUIS are fitted by the following exponentially modified Gaussian function:
\begin{equation}
F(t) = A \cdot e^{\left(\frac{\sigma^2}{2\tau_d^2}-\frac{t-t_0}{2\tau_d}\right)} \cdot \left [1 + erf\left(\frac{t-t_0-\frac{\sigma^2}{2\tau_d}}{\sqrt{2}\sigma}\right)\right]
\end{equation}
where, A is the amplitude of the fit, $t_0$ is time zero as defined by when the sample is excited, $\sigma$ is the width of the Gaussian or the instrument response time, and $\tau_d$ is the decay time constant.

\textbf{Vibrational Density of States Calculations of LLZO}
To find the relaxed structure file inputs for MD calculations, the cubic and tetragonal polymorphs were first modified to reduce electrostatic energy contributions similarly to a procedure in literature.\cite{Mahanta2026} For the initial crystallographic structure files of the cubic (ICSD ID: 238685) cell with Nb doping on the Zr site and the tetragonal (ICSD ID: 14991) cell, clusters of Li\textsuperscript{+} atoms were identified to locate all possible Li\textsuperscript{+} sites, including overlapping sites. Then, the Li\textsuperscript{+} sites were repopulated according to the site occupancy of the polymorph; for c-LLZO, one Li\textsuperscript{+} vacancy was introduced into the cell such that the cell has the stoichiometry \ce{Li_{6.375}La3Zr_{1.5}Nb_{0.5}O12} as opposed to \ce{Li_{6.5}La_{3}Zr_{1.5}Nb_{0.5}O_{12}}. Using the Python Materials Genomics (pymatgen)\cite{Ong2013} and Atomic Simulation Environment (ASE)\cite{Larsen2017} modules, an Ewald electrostatic energy minimization was performed to generate 1000 distinct structures for each polymorph and find the lowest energy configuration. For c-LLZO, only structures with vacancies near Nb dopant atoms were retained to capture local electrostatic interactions. Following the selection of the lowest electrostatic energy structures, variable cell structural relaxation calculations were executed using density functional theory simulations in Quantum ESPRESSO within the projector augmented wave approach.\cite{Giannozzi2009, Giannozzi2017} Calculations for both polymorphs used Perdew-Burke-Ernzerhof (PBE)\cite{Perdew1996} optimized norm conserving Vanderbilt (ONCV)\cite{Hamann2013, Schlipf2015} pseudopotentials, a $2\times2\times2$ uniform k-point grid, and Gaussian smearing of 0.001. Structural relaxations of c-LLZO employed a 200 eV wavefunction cutoff energy while t-LLZO calculations required a 250 eV cutoff. 

Species-projected vibrational DOS calculations using classical MD were computed in the Large-scale Atomic/Molecular Massively Parallel Simulator (LAMMPS).\cite{Thompson2022} Larger simulation cells for MD calculations were found by creating $2\times2\times2$ supercells of the relaxed unit cells, resulting in 1496- and 1536-atom simulation cells for c-LLZO and t-LLZO, respectively. Then, empirical interatomic potential parameters defining a Buckingham potential with a Coulombic pairwise interaction cutoff were applied.\cite{Woodley1999, Bush1994} Following a conjugate gradient energy minimization at 300 K, a 100 ps equilibration in the isothermal-isobaric (NPT) ensemble at 300 K, and a 300 ps equilibration in the canonical ensemble (NVT) at 293 K, simulations in the NVT ensemble at 293 K were conducted for 6 ns for each polymorph. A Nosé-Hoover thermostat and a timestep of 0.5 fs were utilized to sample fast atomic vibrations and resolve high-frequency vibrational modes.\cite{Nose1984, Hoover1985} Finally, the mass-weighted species-projected vibrational DOS were extracted from velocity autocorrelation functions using Fourier transformation. 

\newpage
\begin{acknowledgement}
The authors thank Prof. Kimberly See (Caltech) for use of her lab space for the synthesis and characterization of LLZO. The authors thank Dr. Kunimitsu Kataoka (National Institute of Advanced Industrial Science and Technology) for the c-LLZO single crystals used to synthesize the polycrystalline c-LLZO pellets used in this work. Lastly, the authors thank Jax Dallas (Caltech) for generating and characterizing the THz field used in the presented experiments.
\end{acknowledgement}

\section{Funding}
A.K.L. acknowledges support by the Natural Sciences and Engineering Research Council of Canada, funding reference number 599279. N.A.S. acknowledges support by the National Defense Science and Engineering Graduate Fellowship. This material is based upon work supported by the Air Force Office of Scientific Research under award number FA9550-25-C-B010 in the amount of $\$$144,500. All authors acknowledge support from the U.S. Army DEVCOM ARL Army Research Office (ARO) Energy Sciences Competency, Advanced Energy Materials Program award W911NF-23-1-0001. The equipment used in this work was funded by the Air Force Office of Scientific Research (AFOSR), DURIP grant number FA9550-23-1-0197. 

The views and conclusions contained in this paper are those of the authors and should not be interpreted as representing the official policies, either expressed or implied, of the U.S. Army, U.S. Air Force, or the U.S. Government.

\begin{suppinfo}
X-ray diffraction and Raman characterization of LLZO, time and frequency domain spectra of THz field, ultrafast laser heating LUIS data, long-range antenna pick-up signal.
\end{suppinfo}

\bibliography{references}

\end{document}


\newpage
\section{Supplementary Information}
\renewcommand{\thefigure}{S\arabic{figure}}
\setcounter{figure}{0} 

\begin{figure}[hbt!]
\centering
\includegraphics[width=0.95 \linewidth]{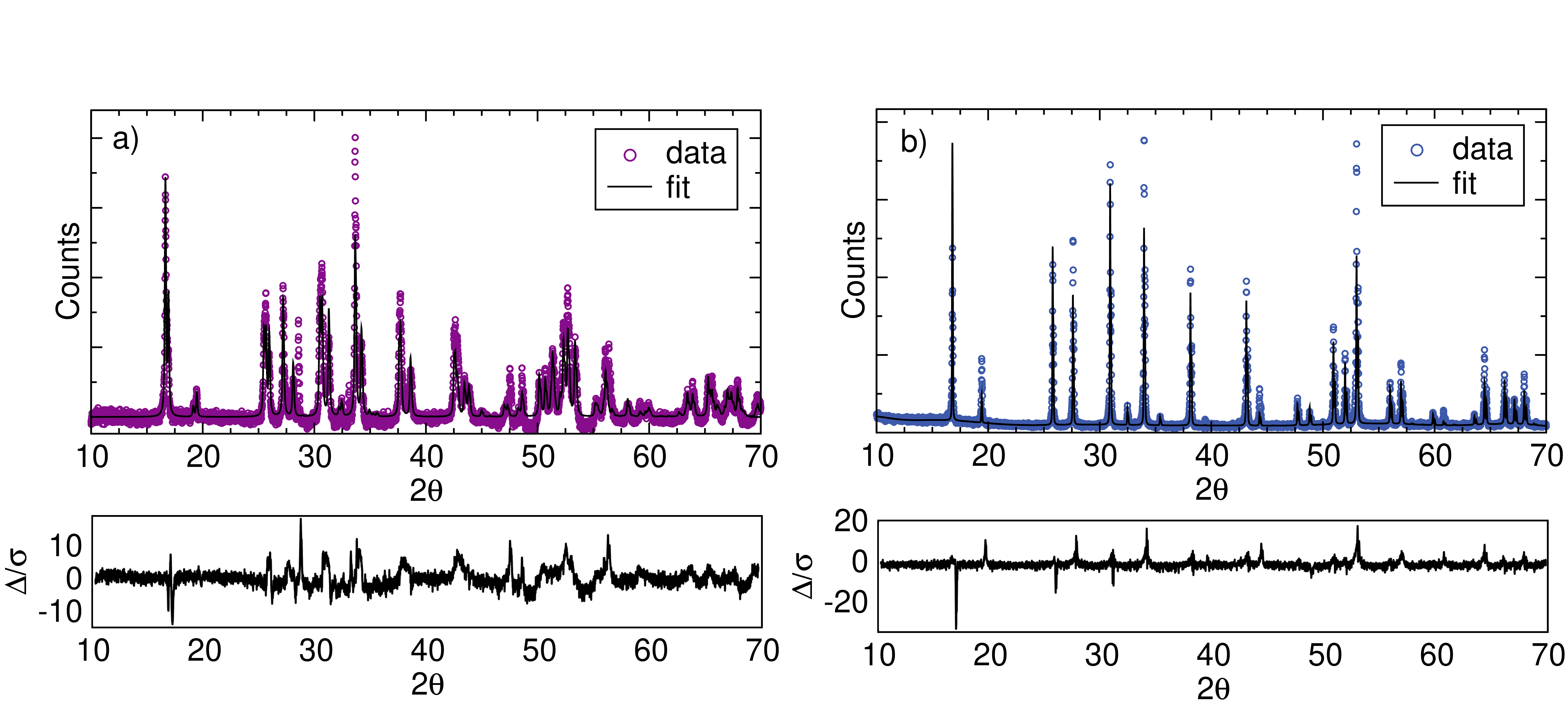}
\caption{\textbf{X-ray diffraction of (a) tetragonal and (b) cubic LLZO.} a) The t-LLZO X-ray diffraction data is fitted to the space group I $4_1$/a c d. b) The Nb-doped c-LLZO X-ray diffraction data is fitted to the space group Ia$\bar3$d.}
\label{si_fig1}
\end{figure}

\begin{figure}[hbt!]
\centering
\includegraphics[width=0.75 \linewidth]{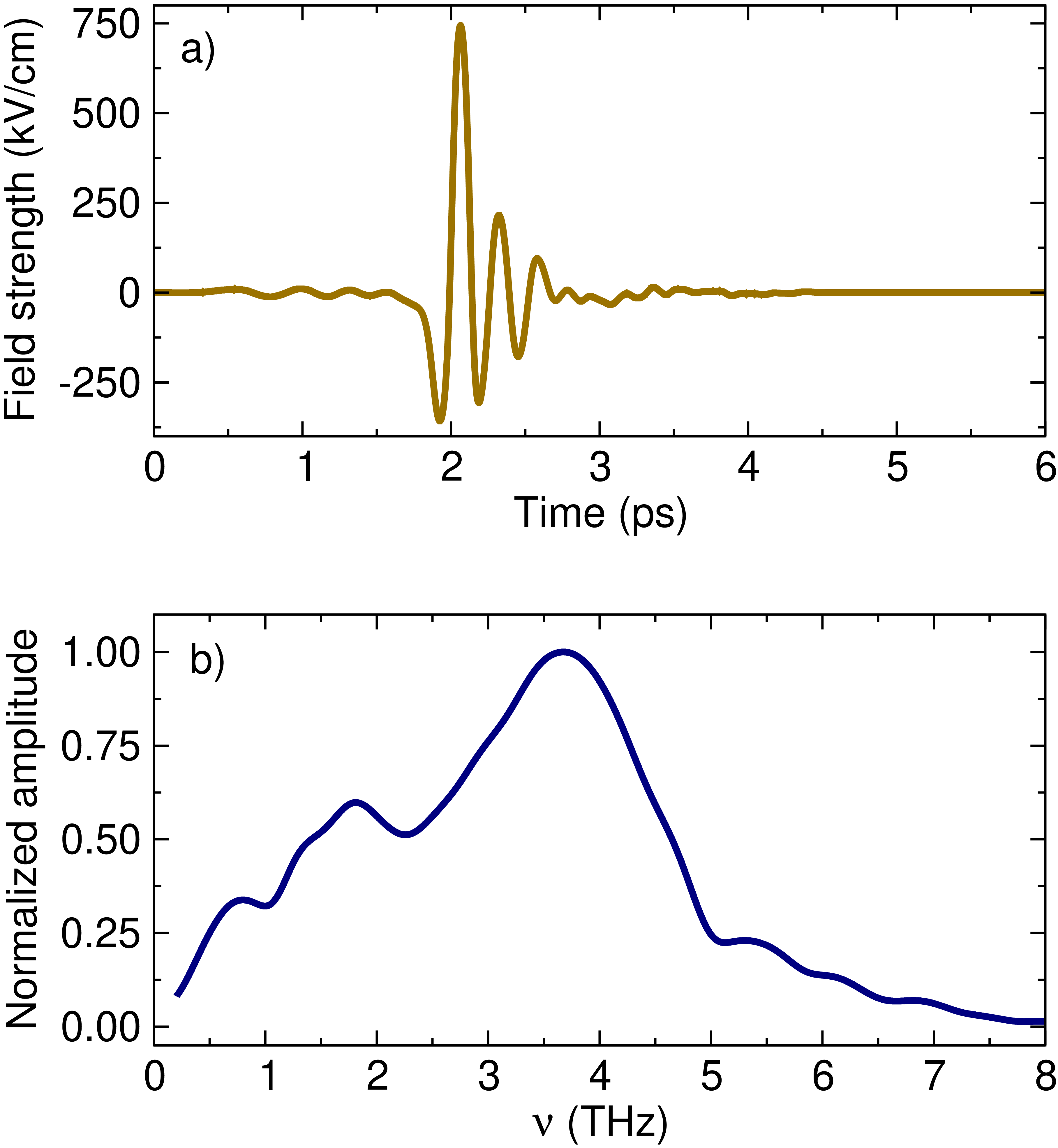}
\caption{\textbf{Time-domain and frequency-domain spectra of THz excitation.} a) The time-domain spectrum (gold) is acquired by electro-optic detection, using a GaP crystal, of the THz field generated by optical rectification in DSTMS. b) A fast Fourier transform is used to obtain the frequency-domain spectrum (blue) of the THz field used to excite both LLZO polymorphs. }
\label{si_fig2}
\end{figure}

\begin{figure}[hbt!]
\centering
\includegraphics[width=0.75 \linewidth]{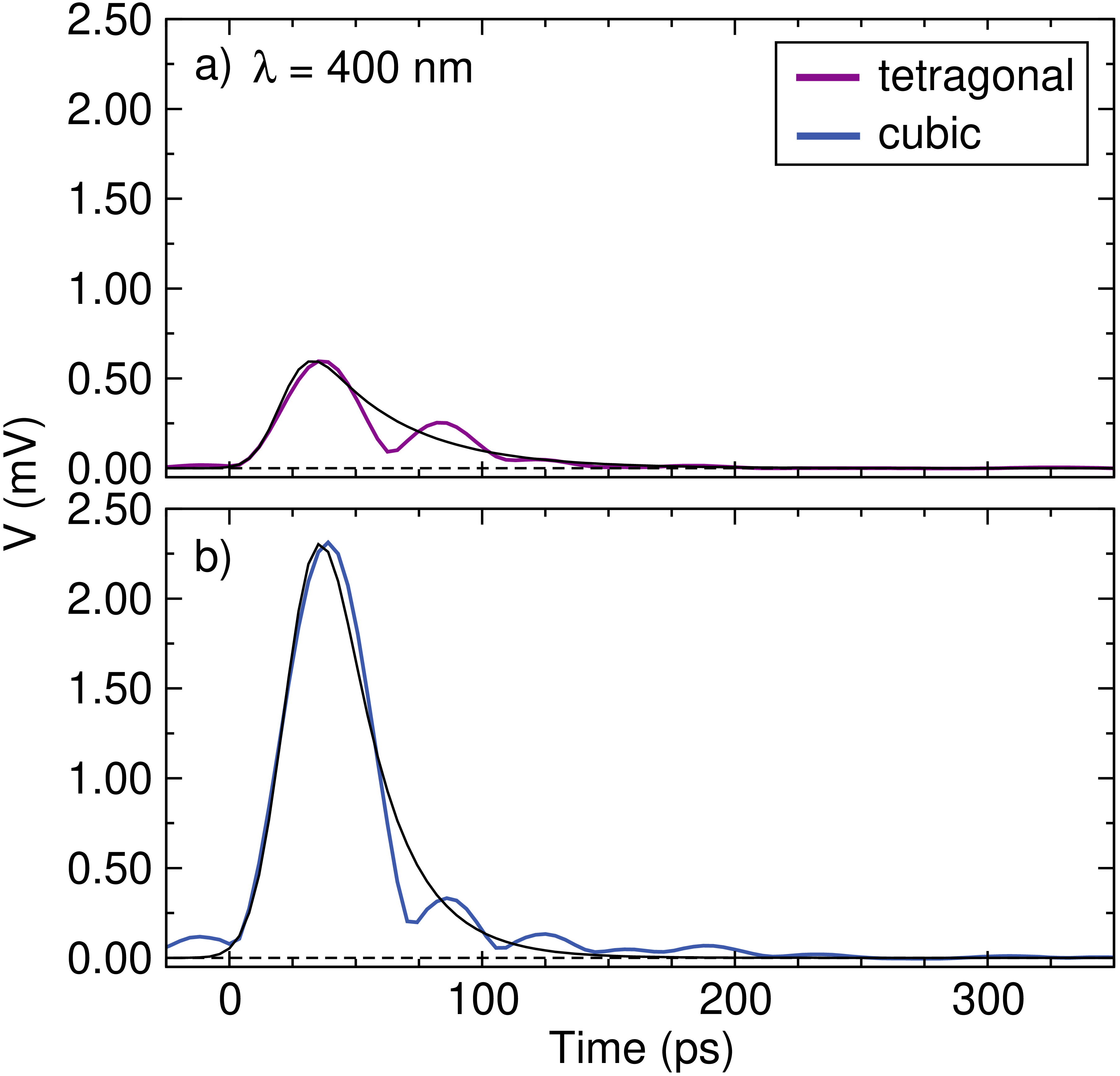}
\caption{\textbf{400-nm-induced perturbations in (a) t-LLZO and (b) c-LLZO impedances.} The rise and decay of a perturbation in the reflected GHz waveform for (a) t-LLZO and (b) c-LLZO representing their change in impedance as a result of a 400 nm excitation ($\sim$12 $\mu$J/pulse) populating defect states. The purple and blue signals are the root-mean-square averages of amplitude-demodulated spectra. The black lines are the exponentially modified Gaussian fit of each spectrum from which the decay time constant is extracted. The decay time constant, reported as a 95 \% confidence interval, for t-LLZO and c-LLZO is 16.7-17.3 picoseconds and 9.83-10.2 picoseconds, respectively. The spectra have been shifted vertically such that the center of the GHz carrier frequency is at zero and horizontally such that time zero represents the 400 nm impulse.}
\label{si_fig3}
\end{figure}

\begin{figure}[hbt!]
\centering
\includegraphics[width=0.75 \linewidth]{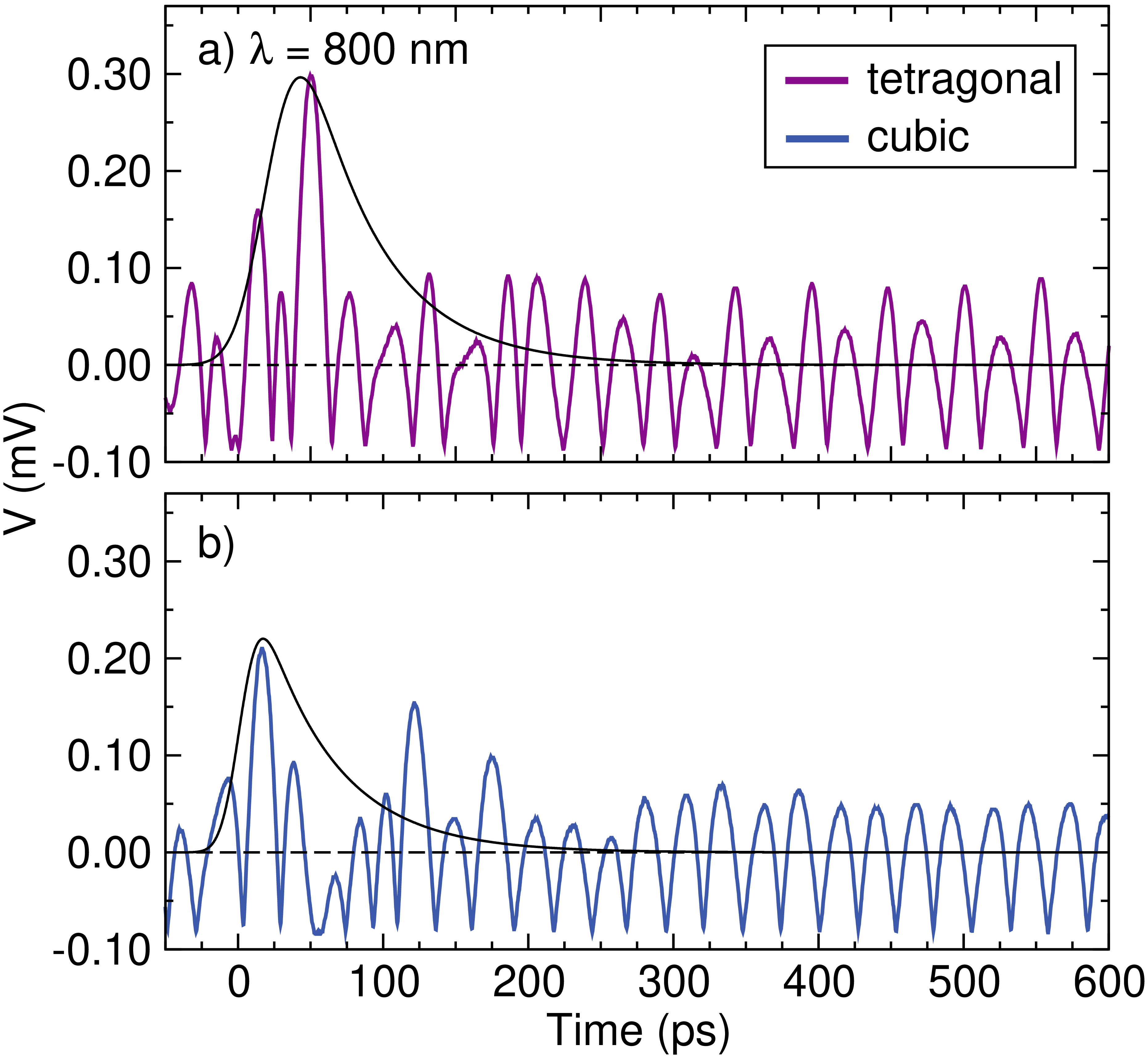}
\caption{\textbf{800-nm-induced perturbations in (a) t-LLZO and (b) c-LLZO impedances.} The rise and decay of a perturbation in the reflected GHz waveform for (a) t-LLZO and (b) c-LLZO representing their change in impedance as a result of a 800 nm excitation ($\sim$8 $\mu$J/pulse) populating defect states. The purple and blue signals are the root-mean-square averages of amplitude-demodulated spectra. The black lines are the exponentially modified Gaussian fit of each spectrum from which the decay time constant is extracted. The decay time constant, reported as a 95 \% confidence interval, for t-LLZO and c-LLZO is 23.1-26.9 picoseconds and 23.0-27.0 picoseconds, respectively. The spectra have been shifted vertically such that the center of the GHz carrier frequency is at zero and horizontally such that time zero represents the 800 nm impulse.}
\label{si_fig4}
\end{figure}

\begin{figure}[hbt!]
\centering
\includegraphics[width=0.75 \linewidth]{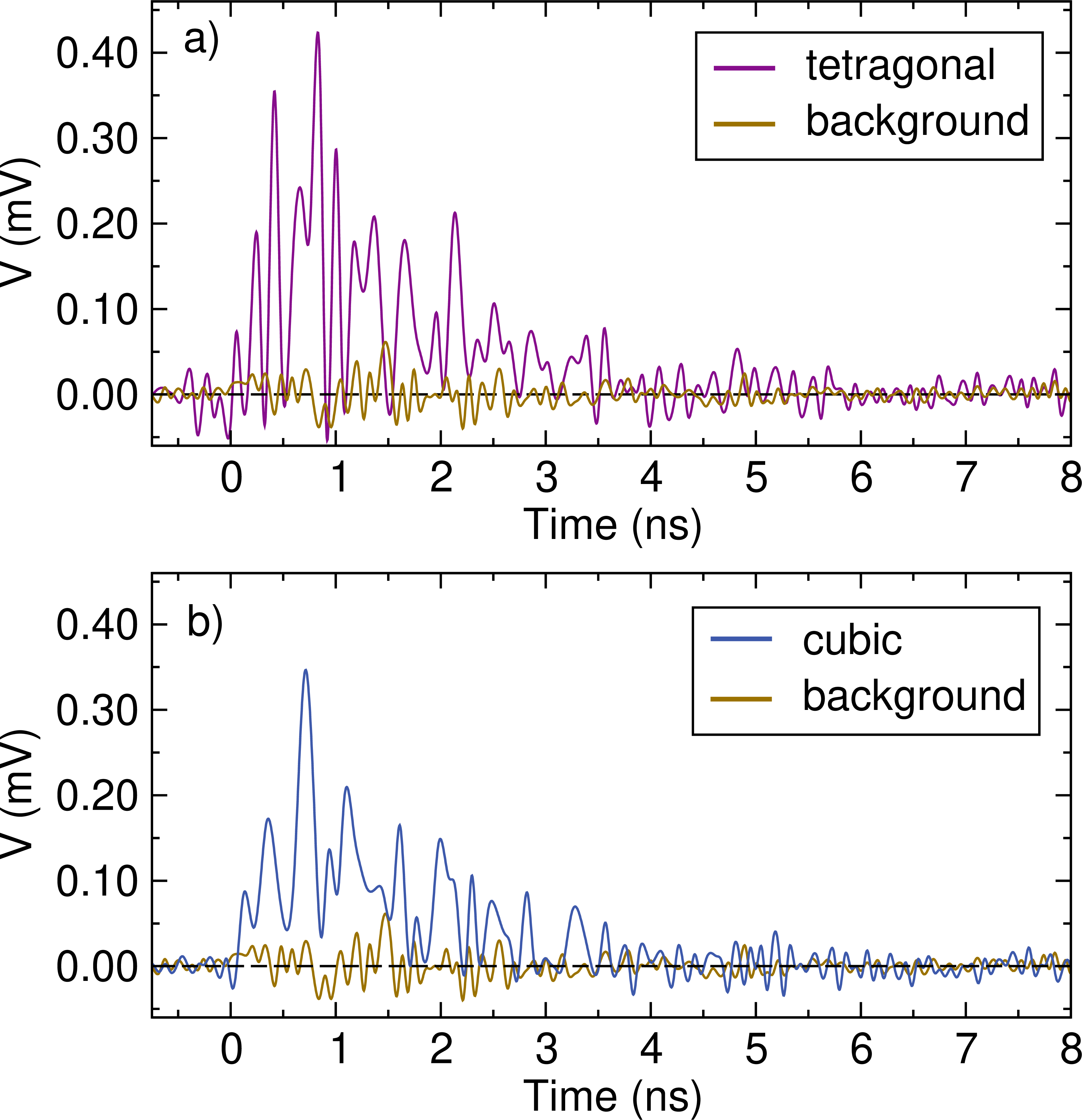}
\caption{\textbf{THz-induced perturbations in (a) t-LLZO and (b) c-LLZO impedances plotted with the long-range antenna pick-up.} The rise and decay of a perturbation in the reflected GHz waveform for (a) t-LLZO and (b) c-LLZO representing their change in impedance as a result of a THz excitation (744 kV/cm) coherently driving phonons in the 0.5-7.5 THz range. The purple and blue signals are the root-mean-square averages of amplitude-demodulated spectra. The spectra have been shifted vertically such that the center of the GHz carrier frequency is at zero and horizontally such that time zero represents the THz impulse. The long-range antenna pick-up, which is present even when the THz is blocked, is shown in gold in both (a) and (b).}
\label{si_fig5}
\end{figure}